\documentclass[12pt]{article}

\def\espaitemps{({\cal V},g)}

\def\be{\begin{equation}}
\def\ee{\end{equation}}
\def\bea{\begin{eqnarray}}
\def\eea{\end{eqnarray}}
\def\bean{\begin{eqnarray*}}
\def\eean{\end{eqnarray*}}

\begin{document}

\title{The universal `energy' operator}
\author{Jos\'e M.M. Senovilla \\
F\'{\i}sica Te\'orica, Universidad del Pa\'{\i}s Vasco, \\
Apartado 644, 48080 Bilbao, Spain \\ 
josemm.senovilla@ehu.es}
\date{}
\maketitle
\begin{abstract} 
The ``positive square" of any tensor is presented in a universal and unified manner, valid in Lorentzian manifolds of arbitrary dimension, and independently of any (anti)-symmetry properties of the tensor. For rank-$m$ tensors, the positive square has rank $2m$. Positive here means {\em future}, that is to say, satisfying the dominant property. The standard energy-momentum and super-energy tensors are recovered as appropriate parts of the general square. A richer structure of principal null directions arises.
\end{abstract} 


\vspace{1cm}
The energy-momentum tensors of a 2-form (such as the electromagnetic field) or of a 1-form (the gradient of the scalar field), or the Bel \cite{Bel1} and Bel-Robinson \cite{Bel} tensors (`super-energy tensor' of the gravitational field, see also \cite{BoS1}), or the super-energy tensors of the electromagnetic \cite{Ch,S} and scalar fields \cite{Tey,S,Tey2}, etc., have either different expressions or different number of indices, or both. A unified, algebraic, perspective of all these tensors was presented in \cite{S} for Lorentzian manifolds of any dimension, where a general method to construct the positive square\footnote{By ``positive square" of a tensor I mean a {\em future} tensor {\em quadratic} in the original one. Future tensors are those satisfying the dominant property, see (\ref{dp}) below; see section 4 in \cite{S}, section 2 in \cite{BS1}, section 2.1 in \cite{GS1}, section 7 in \cite{PP}, and references therein for further details.} of {\em any tensor}, called generically its `superenergy' tensor, was put forward. This method recovered the standard expressions for the energy-momentum and super-energy tensors of the physical fields, keeping therefore its diversity in rank and structure, and allows us to understand that these differences arise because of, and depend on, (i) the number of anti-symmetric sets into which the indices of the fields fall, and (ii) the number of indices within each of these antisymmetric blocks. 

In this short Note, I would like to call attention to a very useful positive square of a tensor that does not seem to have been noticed before. It is the most general ---under some conditions--- future tensor quadratic on the given tensor.

Let $\espaitemps$ be any Lorentzian manifold of arbitrary dimension $n$, with metric tensor $g$ and signature $(-,+,\dots ,+)$.
The basic structure to be used is given by the very simple tensor
\be
G_{\lambda\mu}{}^{\alpha\beta}\equiv \delta_{\lambda}^{\alpha}\delta^{\beta}_{\mu}+
\delta_{\mu}^{\alpha}\delta^{\beta}_{\lambda}-g_{\lambda\mu}g^{\alpha\beta}\label{G}
\ee
which has the properties
\be
G_{\lambda\mu}{}^{\alpha\beta}=G_{(\lambda\mu)}{}^{(\alpha\beta)}=
G^{\alpha\beta}{}_{\lambda\mu},\hspace{2mm}
G^{\rho}{}_{\rho\alpha\beta}=G_{\alpha\beta\rho}{}^{\rho}=
\left(2-n\right)g_{\alpha\beta},\hspace{2mm} G^{\rho}{}_{\alpha\rho\beta}=ng_{\alpha\beta},
\label{prop}
\ee
and when considered as an endomorphism on rank-2 tensors acts as follows:
$$
G_{\lambda\mu}{}^{\alpha\beta}\, t_{\alpha\beta}=2t_{(\lambda\mu)}-t^{\rho}{}_{\rho}\, g_{\lambda\mu}\, .
$$
The `universal (super)-energy operator' is simply given by the $m^{th}$ power of $G$, that is to say, the tensor product $\bigotimes^m G$, where the number of multiplications $m$ is the rank of the tensor to be squared. With indices this operator reads
$$
\fbox{$\displaystyle{
(G\otimes\dots\otimes G)_{\lambda_1\mu_1\dots\lambda_m\mu_m}
{}^{\alpha_1\beta_1\dots\alpha_m\beta_m}\equiv 
G_{\lambda_1\mu_1}{}^{\alpha_1\beta_1}\cdots G_{\lambda_m\mu_m}{}^{\alpha_m\beta_m}=\prod_{i=1}^{m}G_{\lambda_i\mu_i}{}^{\alpha_i\beta_i}
}$}\, .
$$
Then, given {\em any} tensor $A_{\alpha_1\dots\alpha_m}$ one can define its future square by letting this operator act on the tensor product $A\otimes A$
\bea
{\cal E}_{\lambda_1\mu_1\dots\lambda_m\mu_m}\{A\}
\equiv \frac{1}{2}\left(\prod_{i=1}^{m}G_{(\lambda_i\mu_i)}{}^{\alpha_i\beta_i}\right)
A_{\alpha_1\dots\alpha_m}A_{\beta_1\dots\beta_m}=\nonumber \\
=\frac{1}{2} G_{\lambda_1\mu_1}{}^{\alpha_1\beta_1}\cdots G_{\lambda_m\mu_m}{}^{\alpha_m\beta_m}A_{\alpha_1\dots\alpha_m}A_{\beta_1\dots\beta_m}\, .
\label{met}
\eea
Observe that this definition is {\em independent of the dimension $n$}. In order to distinguish this generalization from its predecessors ---the energy-momentum and super-energy tensors--- I propose to call the tensor (\ref{met}) the ``mathematical energy tensor" of $A$.\footnote{Other possible names are the ``causal squares" of $A$ for $\pm {\cal E}\{A\}$, and in particular the ``future square" of $A$ for the plus sign.} Note that one may let the operator act on two {\em different} tensors $A_{\mu_1\dots\mu_m}$ and $B_{\mu_1\dots\mu_m}$ as long as they have the same rank $m$; an equivalent way of defining this operation is (suppressing indices)
$$
{\cal E}\{A,B\}\equiv 2\, {\cal E}\left\{\frac{A+B}{2}\right\}-\frac{1}{2}{\cal E}\{A\}-\frac{1}{2}{\cal E}\{B\}
$$
so that ${\cal E}\{A,B\}={\cal E}\{B,A\}$ and ${\cal E}\{A,A\}={\cal E}\{A\}$.

Now, let me enumerate the more important properties of the tensors (\ref{met}).

{\bf (i)} The mathematical energy tensor of a rank-$m$ tensor has rank $2m$. From (\ref{prop}) and (\ref{met}) one derives
\bean
{\cal E}_{\lambda_1\mu_1\dots\lambda_m\mu_m}\{A,B\}&=&
{\cal E}_{(\lambda_1\mu_1)\dots (\lambda_m\mu_m)}\{A,B\}\, , \\
{\cal E}^{\rho}{}_{\rho\lambda_2\mu_2\dots\lambda_m\mu_m}\{A,B\}&=&\left(2-n\right)
\left(\prod_{i=2}^{m}G_{\lambda_i\mu_i}{}^{\alpha_i\beta_i}\right)
A^{\rho}{}_{\alpha_2\dots\alpha_m}B_{\rho\beta_2\dots\beta_m}
\eean
and similarly for other traces on any pair $\lambda_i\mu_i$.

{\bf (ii)} As is obvious from (\ref{met}), if $A_{\alpha_1\dots\alpha_m}$  is symmetric, or antisymmetric, with respect to two indices $\alpha_i$ and $\alpha_j$, then 
${\cal E}_{\lambda_1\mu_1\dots\lambda_m\mu_m}\{A\}$ is symmetric on the interchange of the corresponding pairs $\lambda_i\mu_i$ and $\lambda_j\mu_j$. Similar properties can be listed for
${\cal E}_{\lambda_1\mu_1\dots\lambda_m\mu_m}\{A,B\}$ if the two tensors have the (anti)-symmetries on corresponding indices. Thus, if 
$A_{\alpha_1\dots\alpha_i\dots\alpha_j\dots\alpha_m}=\epsilon_1 A_{\alpha_1\dots\alpha_j\dots\alpha_i\dots\alpha_m}$ and 
$B_{\alpha_1\dots\alpha_i\dots\alpha_j\dots\alpha_m}=\epsilon_2 B_{\alpha_1\dots\alpha_j\dots\alpha_i\dots\alpha_m}$
were $\epsilon_1^2=\epsilon_2^2=1$, then
$$
{\cal E}_{\lambda_1\mu_1\dots\lambda_i\mu_i\dots\lambda_j\mu_j\dots\lambda_m\mu_m}\{A,B\}=
\epsilon_1\epsilon_2\, 
{\cal E}_{\lambda_1\mu_1\dots\lambda_j\mu_j\dots\lambda_i\mu_i\dots\lambda_m\mu_m}\{A,B\}\, .
$$

{\bf (iii)} Observe that (\ref{met}) coincides with the basic superenergy tensor of any tensor $A_{\alpha_1\dots\alpha_m}$ with {\em no antisymmetries}. To see this, let me recall that the superenergy construction is based on the following fact \cite{S0,S,ES}: there is a precise and unique way to consider {\em any} tensor $A_{\mu_1\dots \mu_m}$ as an {\em $r$-fold form}, that is to say, as an element of $\Lambda^{n_{1}}\otimes\dots\otimes \Lambda^{n_{r}}$ where $\Lambda^p$ is the set of $p$-forms. The well defined natural number $r$ is called the {\em form-structure number} of $A_{\mu_1\dots \mu_m}$ \cite{ES}; obviously, $r\leq m$. Moreover, the set of $r$ natural numbers $n_{1},\dots ,n_{r}$ is {\em univocally} defined, and each $n_\Upsilon$ is called the {\em $\Upsilon$-th block rank}. It is trivial that  $\sum_{\Upsilon =1}^{r}n_{\Upsilon}=m$. Tensors seen in this way are called $r$-fold $(n_{1},\dots ,n_{r})$-forms and denoted by $A_{[n_1]\dots[n_r]}$ \cite{S}.

Given any tensor $A_{\mu_1\dots\mu_m}$ with form structure number $r$, its basic superenergy tensor $T_{\lambda_1\mu_1\dots\lambda_r\mu_r}\{A\}$ has $2r$ indices, it is symmetric on each of the $r$ pairs, and can be defined by \cite{S,S2}:\footnote{Here, $\tilde{A}_{\mu_1\dots \mu_m}$ is the permuted version of $A_{\mu_1\dots \mu_m}$ so that its first $n_1$ indices are those precisely in $[n_1]$, the next $n_2$ indices are those in $[n_2]$, and so on.}
\be
T_{\lambda_1\mu_1\dots\lambda_r\mu_r}\{A\}=\frac{1}{2} 
E_{\lambda_1\mu_1\dots\lambda_r\mu_r}
{}^{\sigma_1\dots\sigma_{n_1}\dots\tau_1\dots\tau_{n_r}}_{\rho_1\dots\rho_{n_1}\dots \nu_1\dots\nu_{n_r}}\, \tilde{A}_{\sigma_1\dots\sigma_{n_1}\dots\tau_1\dots\tau_{n_r}}\tilde{A}^{\rho_1\dots\rho_{n_1}\dots\nu_1\dots\nu_{n_r}} \label{set}
\ee
where
\bean
E_{\lambda_1\mu_1\dots\lambda_r\mu_r}
{}^{\sigma_1\dots\sigma_{n_1}\dots\tau_1\dots\tau_{n_r}}_{\rho_1\dots\rho_{n_1}\dots \nu_1\dots\nu_{n_r}}\equiv \frac{1}{(n_1-1)!} \delta^{\sigma_2\dots\sigma_{n_1}}_{\rho_2\dots\rho_{n_1}}\left(2\delta^{\sigma_1}_{(\lambda_1}g_{\mu_1)\rho_1}-\frac{1}{n_1}\delta^{\sigma_1}_{\rho_1}g_{\lambda_1\mu_1}\right)\times \cdots \nonumber\\
\times \frac{1}{(n_r-1)!} \delta^{\tau_2\dots\tau_{n_r}}_{\nu_2\dots\nu_{n_r}}\left(2\delta^{\tau_1}_{(\lambda_r}g_{\mu_r)\nu_1}-\frac{1}{n_r}\delta^{\tau_1}_{\nu_1}g_{\lambda_r\mu_r}\right)
\eean
and $
\delta^{\mu_1\dots\mu_p}_{\nu_1\dots\nu_p}\equiv p! \delta^{\mu_1}_{[\nu_1}\cdots\delta^{\mu_p}_{\nu_p]}$
is the Kronecker symbol of order $p$. 

It follows that, for the case when $A_{\mu_1\dots \mu_m}$ has no antisymmetries, that is to say, when $r=m$, the two tensors (\ref{set}) and (\ref{met}) coincide:
$$
\mbox{If} \hspace{2mm} r=m\,\,  \Longrightarrow \,\,
{\cal E}_{\lambda_1\mu_1\dots\lambda_m\mu_m}\{A\}=
T_{\lambda_1\mu_1\dots\lambda_m\mu_m}\{A\}\, .
$$
Thus, for those familiar with superenergy tensors, the rule in order to compute (\ref{met}) for general tensors is very simple: {\em just ignore the antisymmetries of the tensor $A$}, its form structure number and block ranks, and build its super-energy tensor as an $m$-fold (1,1,\dots,1)-form.

{\bf (iv)} The most important property is that the mathematical energy tensor (\ref{met}) is a future tensor, see \cite{BS1,GS,GS1}. This means that it always satisfies the {\em dominant property} \cite{S}: for any collection of future-directed vectors $\{\vec{u}_1,\vec{v}_1,\dots ,\vec{u}_m,\vec{v}_m\}$
\be
{\cal E}_{\lambda_1\mu_1\dots\lambda_m\mu_m}\{A\}
u_1^{\lambda_1}v_1^{\mu_1}\cdots u_m^{\lambda_m}v_m^{\mu_m}\geq 0 \, .\label{dp}
\ee
This property permits the finding of estimates on the growth of the tensor field $A$ (if it is subject to some differential field equations) and thereby has been essential in many important applications of the energy-momentum and super-energy tensors, such as the causal propagation of fields \cite{HE,BoS,BS}, the stability of spacetimes \cite{CK,KN}, getting hyperbolizations of many physical fields including gravity \cite{F,F1,AChY,Bo,ChY3,S2}, the existence of global solutions to the gravitational Cauchy problem \cite{KN,FR}, and the propagation of discontinuities \cite{L,S}.

The basic super-energy tensor (\ref{set}) of $A$ is the most general rank-$2r$ tensor ---modulo index permutations, see section 5 in \cite{S}---, symmetric by pairs and quadratic on $A$ satisfying the dominant property. As a matter of fact, the tensor (\ref{met}) is analogously the most general tensor, modulo index permutations, quadratic on $A$ (and without extra $g's$, that is, with rank $2m$) satisfying the dominant property ---and symmetric on each $\lambda_i\mu_i$-pair (see footnote \ref{foot}.) 

The dominant property can be characterized by several different but equivalent conditions, see \cite{S,BS1,GS}. Among them, one can cite (a) that the inequality (\ref{dp}) is strict if all the future-directed vectors $\{\vec{u}_1,\vec{v}_1,\dots ,\vec{u}_m,\vec{v}_m\}$ are {\em timelike} (if $A$ is non-vanishing, of course); (b) that the vectors
$$
{\cal P}^{\alpha}=-{\cal E}^{\alpha}{}_{\mu_1\dots\lambda_m\mu_m}\{A\}v_1^{\mu_1}\dots u_m^{\lambda_m}v_m^{\mu_m}
$$
are future directed for any arbitrary set of future-directed vectors 
$\{\vec{v}_1,\dots ,\vec{u}_m,\vec{v}_m\}$, and analogously by leaving any other single index free; or (iii) that the totally timelike component `dominates' the tensor, meaning that in any orthonormal basis $\{\vec{e}_0,\dots ,\vec{e}_{n-1}\}$ with $\vec{e}_0$ timelike future-directed,
\be
{\cal E}_{00\dots 00}\{A\} \geq |{\cal E}_{\lambda_1\mu_1\dots\lambda_m\mu_m}\{A\} | \label{dp1}
\ee
for all possible values of the indices.

To prove that the dominant property holds for (\ref{met}) one can just use any of the available proofs\footnote{Here is where the symmetrization on each $\lambda_i\mu_i$-pair is needed, as there is no available proof of the dominant property in general dimension without this restriction. However, {\em in four dimensions}, this can be generalized by using spinors and the reasonings in \cite{Ber}. Therefore, it seems plausible that these symmetrizations can be avoided in general dimension. Observe that the superenergy density of the next point (v) is not affected by this at all.\label{foot}} for the superenergy tensors \cite{S,PP}, apply it to an $m$-fold (1,\dots ,1)-form, and note that any possible antisymmetries of the tensor $A$ do not invalidate the proofs. As the property holds for {\em generic} tensors $A$, then nothing changes if the tensor $A$ happens to have some antisymmetries.

{\bf (v)} The previous point concerning the validity of the general proof of the dominant property is actually related to the fact that the generic `electric-magnetic' decomposition of any tensor {\em ignoring} its antisymmetries is essentially the same as the canonical `electric-magnetic' decomposition (see \cite{XYZZ,S1} and section 2 in \cite{S}) in which the form structure number and block ranks play an important role. The reason for this is that the canonical `electric-magnetic' decomposition is nothing but the generic one with some `electric-magnetic' parts ---those corresponding to more than one `electric', or more than one `magnetic', part within one antisymmetric block---vanishing. 

Using this fact, one can easily derive the fact that the totally timelike component arising in (\ref{dp1}) is, up to a multiplicative factor, the so-called superenergy density of the tensor $A$ relative to the unit timelike vector $\vec{e}_0$. This is denoted by $W_{A}(\vec{e}_0)$, it is defined by \cite{S}
$$
W_A\left(\vec{e}_0\right)\equiv
T_{\lambda_1\mu_1\dots\lambda_r\mu_r}\{A\}
e_0^{\lambda_1}e^{\mu_1}_0\dots e_0^{\lambda_r}e_0^{\mu_r} 
$$
and it is equal to half the sum of the positive squares of all the canonical electric-magnetic parts of the tensor $A_{\mu_1\dots\mu_m}$ \cite{S}. The supernergy density is also equal to half the sum of the squares of all the {\em independent} components of $A_{\mu_1\dots\mu_m}$ in any orthonormal basis $\{\vec{e}_{\mu}\}$ containing $\vec{e}_0$ (this is the typical mathematical `energy' of the tensor $A_{\mu_1\dots\mu_m}$):\footnote{There is a typo in Property 3.5 of \cite{S}: a missing factor $\prod_{\Upsilon=1}^{r}\frac{1}{n_{\Upsilon}!}$.}
$$
W_A(\vec{e}_0)=T_{00\dots 0}\{A\}=\frac{1}{2}\left(\prod_{\Upsilon=1}^{r}\frac{1}{n_{\Upsilon}!}\right)
\sum_{\mu_1,\dots,\mu_m=0}^{n-1}
|A_{\mu_1\dots\mu_m}|^2 \, .
$$
It follows then easily that
$$
{\cal E}_{00\dots 00}\{A\} =
\frac{1}{2}\sum_{\mu_1,\dots,\mu_m=0}^{n-1} |A_{\mu_1\dots\mu_m}|^2=
\left(\prod_{\Upsilon=1}^{r}n_{\Upsilon}!\right)W_A(\vec{e}_0)\, .
$$
Therefore, from this point of view, the new tensor (\ref{met}) and the superenergy tensor (\ref{set}) are essentially equivalent.

{\bf (vi)} Point (v) leads to the idea that the superenergy tensor of $A$ is in fact contained as an essential part of the general tensor (\ref{met}), and some symmetric part of the latter is probably nothing but the former supplemented with the necessary $g's$. This is true, and the proof is as follows. Observe that the first summands in the explicit expression of
${\cal E}_{\lambda_1\mu_1\dots\lambda_m\mu_m}\{A\}$ are always of the type $A_{\lambda_1\dots\lambda_m}A_{\mu_1\dots\mu_m}$ and the symmetrizations of this on each $\lambda_i\mu_i$ pair. There are $2^{m-1}$ terms of this type. The next summands are of type
$$
g_{\lambda_1\mu_1}A_{\rho\lambda_2\dots\lambda_m}A^{\rho}{}_{\mu_2\dots\mu_m}
$$
and the symmetrizations of this on each $\lambda_i\mu_i$ pair, together with the similar ones with an explicit $g_{\lambda_i\mu_i}$ and the corresponding contraction on the $i^{th}$ index of $A$. There are a total of $m\, 2^{m-2}$ summands of this type. And so on by extracting explicit metrics and contracting on the corresponding indices of $A\otimes A$. 

Now, suppose that $A_{\lambda_1\dots\lambda_m}$ has some antisymmetries. Then, the symmetric part of the first type of terms is identically zero, and therefore its contribution to the mathematical energy vanishes. This is also the case for the terms of the second type corresponding to a $g_{\lambda_i\mu_i}$ if the indices of $A_{\lambda_1\dots\lambda_m}$ with $\lambda_i$ excluded still have antisymmetries. And so on for the rest of the summands until one arrives at terms with a number of explicit metrics given by $\sum_{\Upsilon =1}^{r} (n_{\Upsilon}-1)$ where $r$ is the form structure number of $A$. Therefore, the following formula hods in general
\bea
{\cal E}_{(\alpha_1\beta_1\dots\alpha_{n_1}\beta_{n_1})}{}^{(\gamma_1\delta_1\dots
\gamma_{n_2}\delta_{n_2})}{}_{\dots\dots(\lambda_1\mu_1\dots\lambda_{n_r}\mu_{n_r})}\{\tilde{A}\}=
\left(\prod_{\Upsilon =1}^{r}(-1)^{n_{\Upsilon}-1}\, n_{\Upsilon}!\right)\times
\nonumber\\
g_{(\alpha_2\beta_2}\cdots g_{\alpha_{n_1}\beta_{n_1}}
T_{\alpha_1\beta_1)}{}^{(\gamma_1\delta_1}{}_{\dots(\lambda_1\mu_1}\{A\}\, g^{\gamma_2\delta_2}\cdots g^{\gamma_{n_2}\delta_{n_2})}\cdots
g_{\lambda_2\mu_2}\cdots g_{\lambda_{n_r}\mu_{n_r})} \, .\label{met-set}
\eea
Thus, the standard basic superenergy tensor is fully contained in (\ref{met}), and can be obtained from (\ref{met}) by means of the important formula (\ref{met-set}). Observe, nevertheless, that the mathematical energy tensor (\ref{met}) has more information than (\ref{set}), as the positive property (\ref{dp}) for {\em different} future-pointing vectors implies.

{\bf (vii)} The {\em principal directions} of a future tensor are defined by the causal vectors $\vec k$ such that the contraction of the tensor on all its indices with $\vec k$ vanishes, see Definition A.2 in \cite{GS}. As remarked above, the inequality (\ref{dp}) is strict if all the vectors are timelike. Therefore, due to the general properties of causal tensors  \cite{BS1,GS}, the principal directions are necessarily null. The principal directions of the tensor (\ref{met}) are thus given in general by the null vectors $\vec k$ such that
\be
k^{\lambda_1}k^{\mu_1}\cdots k^{\lambda_m}k^{\mu_m}
{\cal E}_{\lambda_1\mu_1\dots\lambda_m\mu_m}\{A\}=0.\label{pd}
\ee
Of course, if the tensor $A$ has no antisymmetries, i.e., when $r=m$, then these are simply the principal directions of the superenergy tensor of $A$,
which are in this case the null directions $\vec k$ such that $k^{\lambda_1}\cdots k^{\lambda_m}A_{\lambda_1\dots\lambda_m}=0$.
However, if $r<m$, that is to say, if the tensor $A$ has at least two antisymmetric indices, then {\em every null vector} is a principal direction of ${\cal E}\{A\}$:
$$
\mbox{If} \hspace{2mm} r<m\,\,  \Longrightarrow \,\,
k^{\lambda_1}k^{\mu_1}\cdots k^{\lambda_m}k^{\mu_m}
{\cal E}_{\lambda_1\mu_1\dots\lambda_m\mu_m}\{A\}=0 \,\,\,\,\,\, \forall \vec k \,\,\, (k^{\mu}k_{\mu}=0).
$$
Fortunately, this does not mean that the information on the {\em principal null directions} of $A$ (see e.g. Definition 2 in \cite{PP}), which coincide with the principal directions of its superenergy tensor and are thus characterized by
$$
k^{\lambda_1}k^{\mu_1}\cdots k^{\lambda_r}k^{\mu_r}T_{\lambda_1\mu_1\dots\lambda_r\mu_r}\{A\}=0,
$$
has been lost. This can also be deduced from point (vi). To recover this information, one only has to start suppressing copies of $\vec k$ from formula (\ref{pd}). As a matter of fact, if one does this in order and systematically, {\em the information about the form structure number and block ranks of $A$ can be retrieved}. 

The procedure is based on the fundamental relation (\ref{met-set}) and is as follows. Start by suppressing one copy of $\vec k$, and do this for each $\lambda_i\mu_i$-pair. That is to say, for each $i=1,\dots , m$, compute
$$
k^{\lambda_1}k^{\mu_1}\cdots k^{\lambda_i}\cdots k^{\lambda_m}k^{\mu_m}
{\cal E}_{\lambda_1\mu_1\dots\lambda_i\mu_i\dots\lambda_m\mu_m}\{A\}
$$
(here $k^{\mu_i}$ is absent.)
If for a given $i$ there is at least one null $\vec k$ such that the previous expression is non-zero, then there exists exactly one $j\neq i$ such that the same happens by removing only the $k^{\mu_j}$. In this case the {\em only} antisymmetric indices in $A_{\mu_1\dots\mu_m}$ are the couple $\mu_i$ and $\mu_j$. This is the end of the story in this case, $r=m-1$ and there are $m-1$ blocks with one index and one block with rank 2. If on the contrary, the previous displayed expression vanishes {\em for all} null $\vec k$ and all values of $i=1,\dots ,m$, then either $r<m-1$ or $r=m-1$ and there is one block with rank greater than 2.

The next step, in the latter case, is to remove a second copy of $\vec k$, corresponding to another $\lambda_j\mu_j$-pair, and do this for all possible combinations of $i$ and $j$. If for all $(i,j)$ this vanishes for arbitrary null $\vec k$, then either $r<m-2$ or $r=m-1$ and there is one block with rank greater than 3. In this case, one has to go to the next step by removing one further copy of $\vec k$. If on the contrary, for some pair $(i,j)$ there is a null $\vec k$ such that this does not vanish, then there must exist another $l$ such that the same happens for the pair $(i,l)$. Two possibilities arise here. If also the removal of the pair $k^{\mu_j}k^{\mu_l}$ in (\ref{dp}) gives a non-vanishing expression, then $A_{\mu_1\dots\mu_m}$ is antisymmetric on the indices $\mu_i,\mu_j,\mu_l$ exclusively. Thus, $r=m-1$ and $A$ is a $(m-1)$-fold form with $m-3$ one-index blocks and one block with rank 3. If on the other hand the removal of the pair $k^{\mu_j}k^{\mu_l}$ in (\ref{dp}) vanishes for all null $\vec k$, then there must exist a companion $p$ for $i$ such that the removal of the pair $k^{\mu_i}k^{\mu_p}$ in (\ref{dp}) is a vanishing expression too. Then $A_{\mu_1\dots\mu_m}$ has antisymmetries on exactly the pairs $\mu_i,\mu_p$ and $\mu_j,\mu_l$, and no more antisymmetries. In this case $r=m-2$ and there are $m-4$ one-index blocks and 2 blocks with rank 2. Observe also that, in general, the position of the antisymmetric indices is clearly identified. 

One has to proceed in this manner until the process stops or all the $k^{\mu_i}$ for $i=1,\dots ,m$ have been removed. In general, the principal directions are characterized by the particular null $\vec k$'s which annihilate the first non-generically vanishing expression of type (\ref{dp}) with several $\vec k$-suppressions. As is known, these principal directions can be characterized as the null $\vec k$ such that its inner product followed by the exterior product on {\em all} blocks of $A$ vanishes, see \cite{PP}.

{\bf (viii)} Perhaps one could add to the previous list of properties the manifest and desirable fact that the universal energy operator, or the derived mathematical energy tensor, is very {\em easily remembered}: just use one $G_{\lambda\mu}{}^{\alpha\beta}$ as defined in (\ref{G}) for each pair of corresponding indices of the tensor $A\otimes A$; see (\ref{met}).

Let us consider some illustrative and important examples. 

\paragraph{Rank-2 tensors} Let $A_{\mu\nu}$ be an arbitrary rank-2 tensor. Then  its mathematical energy tensor (\ref{met}) reads
$$
{\cal E}_{\alpha\beta\lambda\mu}\{A\}=
A_{\alpha\lambda}A_{\beta\mu}+A_{\beta\lambda}A_{\alpha\mu}-
g_{\alpha\beta}A_{\rho\lambda}A^{\rho}{}_{\mu}-g_{\lambda\mu}A_{\alpha\rho}A_{\beta}{}^{\rho}+
\frac{1}{2}g_{\alpha\beta}g_{\lambda\mu}A_{\rho\sigma}A^{\rho\sigma}
$$
This can be written independently of any (anti)-symmetry properties of $A$. Suppose now that $A_{\mu\nu}$ happens to be a 2-form $A_{\mu\nu}=A_{[\mu\nu]}\equiv F_{\mu\nu}$, such as the electromagnetic field. Its well-known energy-momentum tensor (which in this case is also the basic superenergy tensor of $F_{\mu\nu}$) is
$$
T_{\lambda\mu}\{F\}=F_{\lambda\rho}F_{\mu}{}^{\rho}-\frac{1}{4}g_{\lambda\mu}F_{\rho\sigma}F^{\rho\sigma}
$$
so that the mathematical energy tensor of $F_{\mu\nu}$ can be rewritten now simply as
$$
{\cal E}_{\alpha\beta\lambda\mu}\{F\}=-2F_{\alpha(\lambda}F_{\mu)\beta}-g_{\alpha\beta}T_{\lambda\mu}\{F\}-g_{\lambda\mu}T_{\alpha\beta}\{F\}\, .
$$
This tensor satisfies the dominant property, as does $T_{\lambda\mu}\{F\}$. Formula (\ref{met-set}) specializes to ${\cal E}_{(\alpha\beta\lambda\mu)}\{F\}=-2g_{(\alpha\beta}T_{\lambda\mu)}\{F\}$. The energy density of $F_{\mu\nu}$ can be equally computed with any of the two tensors due to
$$
{\cal E}_{\alpha\beta\lambda\mu}\{F\}u^{\alpha}u^{\beta}u^{\lambda}u^{\mu}=2\,
T_{\lambda\mu}\{F\}u^{\lambda}u^{\mu}
$$
for any unit timelike vector $\vec u$. The principal null directions of $F_{\mu\nu}$, defined by $k^{\rho}F_{\rho[\mu}k_{\nu]}=0$ ---that is, the null eigenvectors of $F_{\mu\nu}$---, are equally well characterized by any of
$$
{\cal E}_{\alpha\beta\lambda\mu}\{F\}k^{\beta}k^{\lambda}k^{\mu}=0\, \Longleftrightarrow \, T_{\lambda\mu}\{F\}k^{\lambda}k^{\mu}=0.
$$
Observe, however, that the mathematical energy tensor has more information than the energy-momentum tensor. For instance, the pairs of null vectors $\vec k, \vec \ell$ such that
$$
{\cal E}_{\alpha\beta\lambda\mu}\{F\}\ell^{\alpha}\ell^{\beta}k^{\lambda}k^{\mu}=0
$$
define the null directions such that $F_{\mu\nu}\ell^{\mu}k^{\nu}=0$ ---that is, {\em all} timelike 2-planes orthogonal to $F_{\mu\nu}$---, while these cannot be retrieved with $T_{\lambda\mu}\{F\}$.

Recall that any rank-2 tensor can be uniquely decomposed into its symmetric and antisymmetric parts, by means of $A_{\mu\nu}=A_{(\mu\nu)}+A_{[\mu\nu]}\equiv S_{\mu\nu}+F_{\mu\nu}$. Then from the definition of (\ref{met}) one derives
$$
{\cal E}\{A\}={\cal E}\{S\}+{\cal E}\{F\}+2{\cal E}\{S,F\}
$$
which allows us to put in relation the mathematical energy tensor of $A$ with those of its symmetric and antisymmetric parts. Note that this was impossible with the superenergy tensors, as they had different ranks for $S$ and $F$. Observe also that 
${\cal E}_{\alpha\beta\lambda\mu}\{S,F\}=-{\cal E}_{\lambda\mu\alpha\beta}\{S,F\}$
so that the (super)-energy density of $A$ is the sum of those of $S$ and $F$.

\paragraph{Rank-3 tensors} Let $A_{\beta\mu\nu}$ be an arbitrary rank-3 tensor. Then its mathematical energy tensor (\ref{met}) reads
\bean
{\cal E}_{\alpha\beta\lambda\mu\tau\nu}\{A\}=
A_{\alpha\lambda\tau}A_{\beta\mu\nu}+A_{\beta\lambda\tau}A_{\alpha\mu\nu}+
A_{\alpha\mu\tau}A_{\beta\lambda\nu}+A_{\beta\mu\tau}A_{\alpha\lambda\nu}\hspace{1cm}\\
-g_{\alpha\beta}\left(A_{\rho\lambda\tau}A^{\rho}{}_{\mu\nu}+
A_{\rho\mu\tau}A^{\rho}{}_{\lambda\nu}\right)
-g_{\lambda\mu}\left(A_{\alpha\rho\tau}A_{\beta}{}^{\rho}{}_{\nu}+
A_{\beta\rho\tau}A_{\alpha}{}^{\rho}{}_{\nu}\right)-g_{\tau\nu}\left(
A_{\alpha\lambda\rho}A_{\beta\mu}{}^{\rho}+A_{\beta\lambda\rho}A_{\alpha\mu}{}^{\rho}\right)\\
+g_{\alpha\beta}g_{\lambda\mu}A_{\rho\sigma\tau}A^{\rho\sigma}{}_{\nu}+
g_{\alpha\beta}g_{\tau\nu}A_{\rho\lambda\sigma}A^{\rho}{}_{\mu}{}^{\sigma}+
g_{\lambda\mu}g_{\tau\nu}A_{\alpha\rho\sigma}A_{\beta}{}^{\rho\sigma}
-\frac{1}{2}g_{\alpha\beta}g_{\lambda\mu}g_{\tau\nu}A_{\rho\sigma\zeta}A^{\rho\sigma\zeta}\, .
\eean
Suppose now that $A_{\beta\mu\nu}$ happens to be antisymmetric on (say) the last two indices, so that $A$ is in fact a double (1,2)-form $A_{\beta\mu\nu}=A_{\beta[\mu\nu]}\equiv L_{\beta\mu\nu}$; this is the case of the covariant derivative of the electromagnetic field, $\nabla_{\beta}F_{\mu\nu}$. The basic superenergy tensor of $L$ is \cite{S}
$$
T_{\alpha\beta\lambda\mu}\{L\}=L_{\alpha\lambda\rho}L_{\beta\mu}{}^{\rho}+
L_{\beta\lambda\rho}L_{\alpha\mu}{}^{\rho}-g_{\alpha\beta}L_{\rho\lambda\sigma}L^{\rho}{}_{\mu}{}^{\sigma}-\frac{1}{2}
g_{\lambda\mu}L_{\alpha\rho\sigma}L_{\beta}{}^{\rho\sigma}+\frac{1}{4}
g_{\alpha\beta}g_{\lambda\mu}L_{\rho\sigma\zeta}L^{\rho\sigma\zeta}
$$
and thus, the mathematical energy tensor of $L_{\beta\mu\nu}$ can be rewritten as
\bean
{\cal E}_{\alpha\beta\lambda\mu\tau\nu}\{L\}= -g_{\tau\nu}T_{\alpha\beta\lambda\mu}\{L\}
-g_{\lambda\mu}T_{\alpha\beta\tau\nu}\{L\}+L_{\alpha\lambda\tau}L_{\beta\mu\nu} \\
+L_{\beta\lambda\tau}L_{\alpha\mu\nu}+
L_{\alpha\mu\tau}L_{\beta\lambda\nu}+L_{\beta\mu\tau}L_{\alpha\lambda\nu}-g_{\alpha\beta}\left(L_{\rho\lambda\tau}L^{\rho}{}_{\mu\nu}+
L_{\rho\mu\tau}L^{\rho}{}_{\lambda\nu}\right)\, .
\eean
Formula (\ref{met-set}) specializes to 
${\cal E}_{\alpha\beta(\lambda\mu\tau\nu)}\{L\}=-2T_{\alpha\beta(\lambda\mu}\{L\}g_{\tau\nu)}$. 

Consider finally the case where $A_{\beta\mu\nu}$ is a 3-form, that is to say, $A_{\beta\mu\nu}=A_{[\beta\mu\nu]}\equiv H_{\beta\mu\nu}$. The superenergy tensor of $H$ (also called its energy-momentum tensor in this case) is
$$
T_{\lambda\mu}\{H\}=\frac{1}{2}\left(H_{\lambda\rho\sigma}H_{\mu}{}^{\rho\sigma}-\frac{1}{6}g_{\lambda\mu}H_{\rho\sigma\zeta}H^{\rho\sigma\zeta}\right)
$$
hence
\bean
{\cal E}_{\alpha\beta\lambda\mu\tau\nu}\{H\}=
H_{\alpha\lambda\tau}H_{\beta\mu\nu}+H_{\beta\lambda\tau}H_{\alpha\mu\nu}+
H_{\alpha\mu\tau}H_{\beta\lambda\nu}+H_{\beta\mu\tau}H_{\alpha\lambda\nu} \hspace{1cm} \\
-g_{\alpha\beta}\left(H_{\rho\lambda\tau}H^{\rho}{}_{\mu\nu}+
H_{\rho\mu\tau}H^{\rho}{}_{\lambda\nu}\right)
-g_{\lambda\mu}\left(H_{\alpha\rho\tau}H_{\beta}{}^{\rho}{}_{\nu}+
H_{\beta\rho\tau}H_{\alpha}{}^{\rho}{}_{\nu}\right)\hspace{1cm} \\-g_{\tau\nu}\left(
H_{\alpha\lambda\rho}H_{\beta\mu}{}^{\rho}+H_{\beta\lambda\rho}H_{\alpha\mu}{}^{\rho}\right)
+2g_{\alpha\beta}g_{\lambda\mu}T_{\tau\nu}\{H\}+
2g_{\alpha\beta}g_{\tau\nu}T_{\lambda\mu}\{H\}+
2g_{\lambda\mu}g_{\tau\nu}T_{\alpha\beta}\{H\}
\eean
and (\ref{met-set}) reduces now to 
${\cal E}_{(\alpha\beta\lambda\mu\tau\nu)}\{H\}=6g_{(\alpha\beta}g_{\lambda\mu}T_{\tau\nu)}\{H\}$.

Comments analogous to that of rank-2 tensors concerning superenergy density and principal null directions are in order for all cases here. I omit them for the sake of brevity. 

\paragraph{Double (2,2)-forms} One can do the same as before for rank-4 tensors, but in order to keep within a reasonable length let me just consider the important case of double (2,2)-forms, which include the Riemann and Weyl tensors, as well as the tensor product of the electromagnetic field with itself. Let $A_{\alpha\beta\lambda\mu}=A_{[\alpha\beta][\lambda\mu]}$ by any double (2,2)-form, not necessarily symmetric in the interchange of the antisymmetric pairs. Its mathematical energy tensor is
\bean
{\cal E}_{\alpha\beta\gamma\delta\lambda\mu\tau\nu}\{A\}=
A_{\alpha\gamma\lambda\tau}A_{\beta\delta\mu\nu}+
A_{\beta\gamma\lambda\tau}A_{\alpha\delta\mu\nu}+
A_{\alpha\delta\lambda\tau}A_{\beta\gamma\mu\nu}+
A_{\beta\delta\lambda\tau}A_{\alpha\gamma\mu\nu}\\
+A_{\alpha\gamma\mu\tau}A_{\beta\delta\lambda\nu}+
A_{\beta\gamma\mu\tau}A_{\alpha\delta\lambda\nu}+
A_{\alpha\delta\mu\tau}A_{\beta\gamma\lambda\nu}+
A_{\beta\delta\mu\tau}A_{\alpha\gamma\lambda\nu}\\
-g_{\alpha\beta}\left( A_{\rho\gamma\lambda\tau}A^{\rho}{}_{\delta\mu\nu}+
A_{\rho\delta\lambda\tau}A^{\rho}{}_{\gamma\mu\nu}
+A_{\rho\gamma\mu\tau}A^{\rho}{}_{\delta\lambda\nu}+
A_{\rho\delta\mu\tau}A^{\rho}{}_{\gamma\lambda\nu}\right)\\
-g_{\gamma\delta}\left(A_{\alpha\rho\lambda\tau}A_{\beta}{}^{\rho}{}_{\mu\nu}+
A_{\beta\rho\lambda\tau}A_{\alpha}{}^{\rho}{}_{\mu\nu}
+A_{\alpha\rho\mu\tau}A_{\beta}{}^{\rho}{}_{\lambda\nu}+
A_{\beta\rho\mu\tau}A_{\alpha}{}^{\rho}{}_{\lambda\nu}\right)\\
-g_{\lambda\mu}\left(A_{\alpha\gamma\rho\tau}A_{\beta\delta}{}^{\rho}{}_{\nu}+
A_{\beta\gamma\rho\tau}A_{\alpha\delta}{}^{\rho}{}_{\nu}+
A_{\alpha\delta\rho\tau}A_{\beta\gamma}{}^{\rho}{}_{\nu}+
A_{\beta\delta\rho\tau}A_{\alpha\gamma}{}^{\rho}{}_{\nu}\right)\\
-g_{\tau\nu}\left(A_{\alpha\gamma\lambda\rho}A_{\beta\delta\mu}{}^{\rho}+
A_{\beta\gamma\lambda\rho}A_{\alpha\delta\mu}{}^{\rho}+
A_{\alpha\delta\lambda\rho}A_{\beta\gamma\mu}{}^{\rho}+
A_{\beta\delta\lambda\rho}A_{\alpha\gamma\mu}{}^{\rho}\right)\\
+g_{\alpha\beta}g_{\gamma\delta}\left(A_{\rho\sigma\lambda\tau}A^{\rho\sigma}{}_{\mu\nu}+
A_{\rho\sigma\mu\tau}A^{\rho\sigma}{}_{\lambda\nu}\right)+
g_{\lambda\mu}g_{\tau\nu}\left(A_{\alpha\gamma\rho\sigma}A_{\beta\delta}{}^{\rho\sigma}+
A_{\beta\gamma\rho\sigma}A_{\alpha\delta}{}^{\rho\sigma}\right)\\
+g_{\gamma\delta}g_{\tau\nu}T_{\alpha\beta\lambda\mu}\{A\}+
g_{\gamma\delta}g_{\lambda\mu}T_{\alpha\beta\tau\nu}\{A\}+
g_{\alpha\beta}g_{\lambda\mu}T_{\gamma\delta\tau\nu}\{A\}+
g_{\alpha\beta}g_{\tau\nu}T_{\gamma\delta\lambda\mu}\{A\}
\eean
where
\bean
T_{\alpha\beta\lambda\mu}\{A\}=
A_{\alpha\rho\lambda\sigma}
A_{\beta}{}^{\rho}{}_{\mu}{}^{\sigma}
+A_{\alpha\rho\mu\sigma}
A_{\beta}{}^{\rho}{}_{\lambda}{}^{\sigma}-\hspace{1cm}\nonumber\\
-\frac{1}{2}g_{\alpha\beta}
A_{\rho\tau\lambda\sigma}A^{\rho\tau}{}_{\mu}{}^{\sigma}
-\frac{1}{2}g_{\lambda\mu}
A_{\alpha\rho\sigma\tau}A_{\beta}{}^{\rho\sigma\tau}+
\frac{1}{8}g_{\alpha\beta}g_{\lambda\mu}
A_{\rho\tau\sigma\nu}
A^{\rho\tau\sigma\nu}
\eean
is the basic superenergy tensor of $A$, also called its Bel tensor \cite{Bel1,BoS1,S0,S}. (The mathematical energy tensor of any rank-4 tensor can be easily read off from the above formulas.) Formula (\ref{met-set}) becomes now
$$
{\cal E}_{(\alpha\beta\gamma\delta)(\lambda\mu\tau\nu)}\{A\}=
2g_{(\alpha\beta}T_{\gamma\delta)(\lambda\mu}\{A\}g_{\tau\nu)}\, .
$$
The principal null directions of $A_{\alpha\beta\lambda\mu}$, defined by $k^{\rho}k^{\sigma}k_{[\beta}A_{\alpha]\rho\sigma[\mu}k_{\nu]}=0$  \cite{PP}, can be characterized by either $T_{\alpha\beta\lambda\mu}\{A\}k^{\alpha}k^{\beta}k^{\lambda}k^{\mu}=0$ or, equivalently, by
$$
{\cal E}_{\alpha\beta\gamma\delta\lambda\mu\tau\nu}\{A\}
k^{\alpha}k^{\beta}k^{\gamma}k^{\lambda}k^{\mu}k^{\tau}=0\, .
$$
As remarked before, however, the mathematical energy tensor has more information than the superenergy tensor. For instance, the quadruplets of null vectors $\vec k, \vec \ell, \vec n,\vec m$ such that
$$
{\cal E}_{\alpha\beta\gamma\delta\lambda\mu\tau\nu}\{A\}
\ell^{\alpha}\ell^{\beta}n^{\gamma}n^{\delta}k^{\lambda}k^{\mu}m^{\tau}m^{\nu}=0
$$
provide a richer geometrical structure for double (2,2)-forms. 

To finalize, recall that the Riemann tensor $R_{\alpha\beta\lambda\mu}$ of the Lorentzian manifold can be decomposed into the sum of the Weyl conformal tensor $C_{\alpha\beta\lambda\mu}$, which is trace free, and a second part $E_{\alpha\beta\lambda\mu}$ depending on the Ricci tensor exclusively, see e.g. \cite{HE,BoS1,S}. This decomposition implies immediately
$$
{\cal E}\{R\}={\cal E}\{C\}+{\cal E}\{E\} + 2{\cal E}\{C,E\}.
$$
which is a generic decomposition implying that of the Bel tensor into the Bel-Robinson and the so-called `pure matter' and `matter-gravity coupling' super-energy tensors \cite{BoS1,S}.

All in all, it seems that the definition (\ref{met}) is well-posed and mathematically sound, recovers all good properties of previous super-energy and energy-momentum tensors, is generic, valid in arbitrary dimension and for any tensor fields irrespective of their (anti)-symmetry properties, easy to use, provides more information concerning the seed tensor, is potentially fruitful because many properties can be proved by using the generic formula (\ref{met}) and the properties of $G_{\alpha\beta}{}^{\lambda\mu}$ in (\ref{G}) rather than the cumbersome explicit formulas for super-energy tensors, and last but not least, it is easily remembered.

\section*{Acknowledgements}
I thank G. Bergqvist for very interesting remarks, specially concerning footnote \ref{foot}. Financial support under
grants FIS2004-01626 of the Spanish CICyT and 
no. 9/UPV 00172.310-14456/2002 of the University of the Basque 
Country is acknowledged.

\end{document}